\documentclass[a4paper,preprintnumbers,floatfix,superscriptaddress,pra,twocolumn,showpacs,longbibliography]{revtex4-1}

\usepackage{amsmath, amsthm, amssymb,amsfonts}
\usepackage{graphicx}
\usepackage{dcolumn}
\usepackage{bm}
\usepackage{bbm}
\usepackage{hyperref}
\usepackage{mathtools}
\usepackage{comment}
\usepackage{color}
\usepackage{multirow}

\newcommand{\ket}[1]{| #1 \rangle}

\newcommand{\mean}[1]{\left\langle #1 \right\rangle}

\newcommand{\figref}[1]{Fig.~\ref{#1}}
\newcommand{\tabref}[1]{Table~\ref{#1}}
\newcommand{\secref}[1]{Sec.~\ref{#1}}

\DeclareGraphicsExtensions{.pdf,.png,.jpg}

\makeindex

\begin{document}
\title{Bell scenarios with communication}
\date{\today}

\author{J.~B.\ Brask}
\affiliation{Group of Applied Physics, University of Geneva, 1211 Geneva, Switzerland}
\author{R.\ Chaves}
\affiliation{International Institute of Physics, Federal University of Rio Grande do Norte, 59070-405 Natal, Brazil}
\affiliation{Institute for Theoretical Physics, University of Cologne, 50937 Cologne, Germany}

\begin{abstract}
Classical and quantum physics provide fundamentally different predictions about experiments with separate observers that do not communicate, a phenomenon known as quantum nonlocality. This insight is a key element of our present understanding of quantum physics, and also enables a number of information processing protocols with security beyond what is classically attainable. Relaxing the pivotal assumption of no communication leads to new insights into the nature quantum correlations, and may enable new applications where security can be established under less strict assumptions. Here, we study such relaxations where different forms of communication are allowed. We consider communication of inputs, outputs, and of a message between the parties. Using several measures, we study how much communication is required for classical models to reproduce quantum or general no-signalling correlations, as well as how quantum models can be augmented with classical communication to reproduce no-signalling correlations.
\end{abstract}

\maketitle

\section{Introduction}

As realised by Bell \cite{Bell1964}, classical and quantum physics provide fundamentally different predictions about experiments with separate observers that do not communicate. Measurements on entangled quantum states shared between the parties can display correlations which are not captured by any classical description. That is, any description in which observations are described in terms of causal relations between classical random variables and which stipulates that the actions of one party cannot influence the local observations of other, separate parties. Such local causal descriptions imply restrictions on the experimental data -- the famous Bell inequalities -- which can be violated if the data is obtained via local measurements by each observer on shared, entangled quantum states. This phenomenon is known as quantum nonlocality, or simply nonlocality.

Today, nonlocality forms a cornerstone of our understanding of quantum theory \cite{Brunner2014}, and at the same time is the enabling resource for a number of information processing protocols. Nonlocality was demonstrated definitively in several recent experiments \cite{Hensen2015,Shalm2015,Giustina2015}. This paves the way for applications in ultra-secure cryptographic protocols \cite{Acin2006}, communication complexity \cite{Buhrman2010}, randomness certification \cite{Pironio2010,Colbeck2009}, and amplification \cite{Colbeck2012}.

The assumption of no communication is crucial for nonlocality. Clearly, if arbitrary communication is allowed between the parties, any correlations can be explained classically, and there is no nonlocality. In the context of applications, under the assumption of no communication, the observation of nonlocality certifies the non-classical nature of the data, which is an important step for establishing security. Given this importance, it is natural to ask how robust nonlocality is to relaxations of this assumption. How much communication is required to restore a classical explanation? Because of the many possible scenarios with varying numbers of observers and forms of communication, this question has a very rich structure and can be tackled from many different complementary perspectives.

Since entanglement is a prerequisite for nonlocality, one approach is to understand how much communication is needed to simulate all possible correlations arising from given entangled states. With two parties, Toner and Bacon proved that, somewhat unexpectedly, a single bit of classical communication is enough to simulate all correlations obtained by dichotomic, projective measurements on a maximally entangled state of qubits \cite{Toner2003}. For maximally entangled states of arbitrary dimension, two bits of communication are sufficient \cite{Regev2009} and necessary \cite{Vertesi2009} to reproduce all correlations from dichotomic measurements. Multipartite generalizations are also known, for example for the Greenberger-Horne-Zeilinger states \cite{Broadbent2009,Branciard2011}.

Another approach is to focus on correlations arising from a specific number of measurements with a fixed number of outcomes, corresponding to a particular set of Bell inequalities. Simulating the correlations in such a fixed Bell scenario is less demanding than the simulation of all possible correlations from an entangled state. Several complementary avenues have been pursued. One can derive Bell inequalities where a given amount of communication is allowed between the parties \cite{Bacon2003,Maxwell2014}. Alternatively, given nonlocal correlations, one may obtain the minimal average communication necessary to reproduce them via classical resources \cite{Pironio2003}. These results consider that one or more parties communicate information about their measurement choices via a message of limited dimension (smaller than the number of measurements, otherwise the problem is trivial \cite{Bacon2003}). Other communication scenarios are also possible, as noted recently \cite{Pawlowski2010NJP,Chaves2015,Ringbauer2016}, where for instance the measurement outcomes are communicated. Furthermore, instead of average communication, different figures of merit may in some cases be more appropriate. For example, measures of causal influence \cite{Janzing2013,Chaves2015} or the message entropy \cite{Chaves2015entropy}.

In this paper, our aim is to further develop the framework for the study of Bell scenarios with communication. In particular, we consider four different scenarios: \emph{(i)} direct causal influence from the measurement choice (input) of one party on the outcome (output) of another, \emph{(ii)} direct causal influence from the output of one party on the output of another, \emph{(iii)} communication of information about the inputs via a message, and \emph{(iv)} quantum correlations augmented with limited communication. We focus on bipartite settings with two observers.

For scenario (i), we extend the results of \cite{Chaves2015} for the minimum amount of causal influence required to simulate given nonlocal correlations. For scenario (ii), we prove that whenever the number of outputs for one party is at least as large as the number of inputs, communication of the outputs is enough to simulate all non-signalling correlations with uniform marginals. We further show that this is not generally true for non-uniform marginals. For settings where the number of outputs is smaller than the number of inputs, we conjecture a general class of Bell inequalities which can be violated by quantum mechanics. In scenario (iii), we obtain a new family of Bell inequalities valid for an arbitrary number of inputs and varying amounts of communication. Considering fixed numbers of inputs and outputs, we also compute the minimum communication entropy required to achieve a given violation of certain Bell inequalities. Finally, in scenario (iv) we show how quantum correlations augmented with limited communication can be used to simulate no-signalling postquantum correlations.

The paper is organized as follows. In \secref{sec.scenarios} we describe the scenarios we consider and the concepts and tools used in the rest of the paper. In Secs.~\ref{sec.locrelax}, \ref{sec.outcomes}, \ref{sec.message}, and \ref{sec.augmenting} we analyse scenarios (i), (ii), (iii) and (iv) in detail, respectively. Finally, in \secref{sec.discussion} we conclude with a summary and discussion of our results and point out open questions that we believe deserve further investigation.

\section{Scenarios, concepts and tools}
\label{sec.scenarios}

In the following we define the notation and mathematical tools that we will use, and give a precise description of the causal models corresponding to the scenarios we will study. The basic framework we will need is Bayesian networks and their graphical representation in terms of directed acyclic graphs (DAGs) \cite{Pearlbook}. We also define the notions of classical, quantum and no-signalling correlations, and define measures of relaxations of locality.

\begin{figure}[t]
\includegraphics[width=0.99\columnwidth]{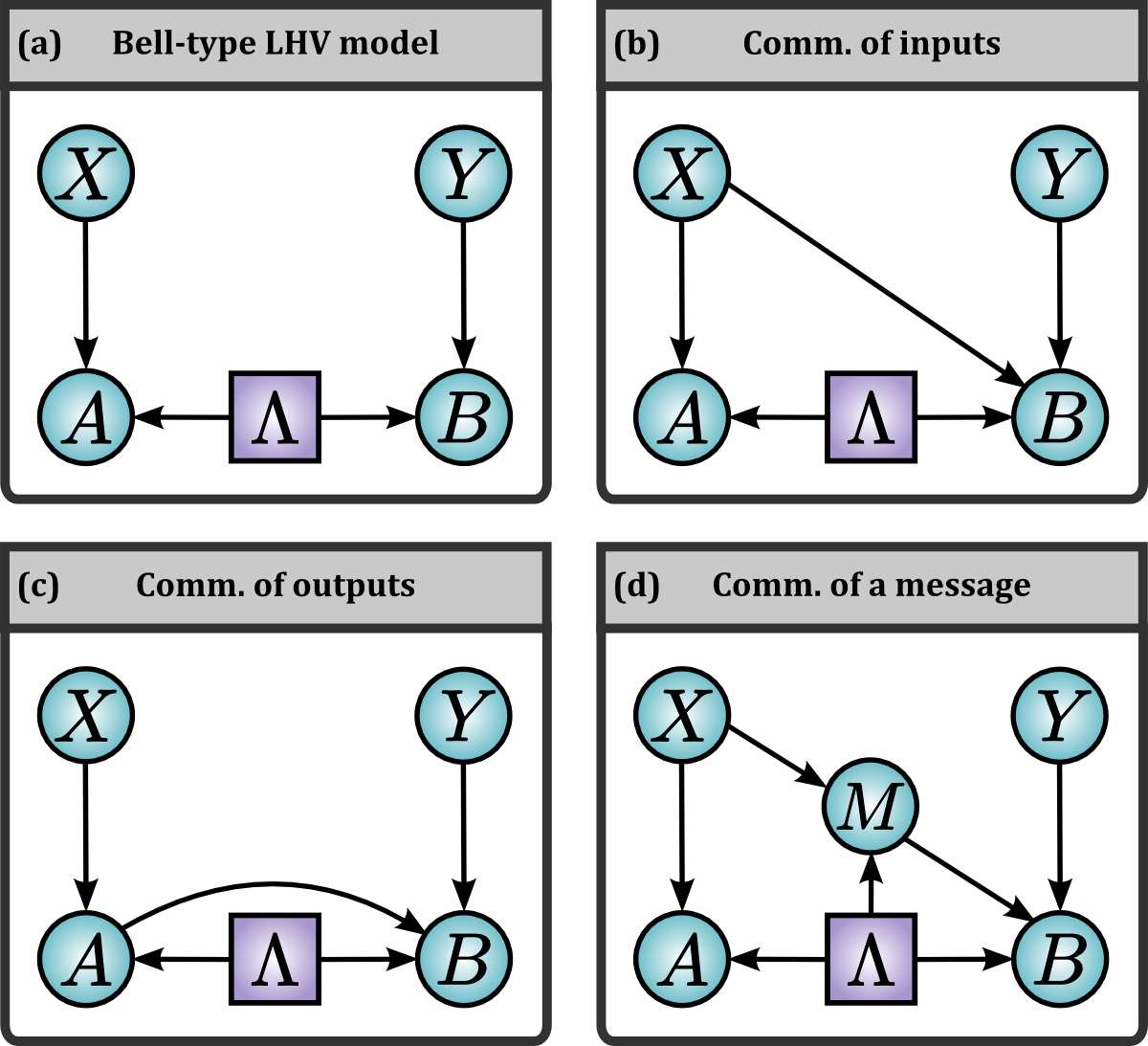}
\caption{DAGs for the causal models considered in the text. \textbf{(a)} Local-hidden-variable model for a bipartite Bell scenario. \textbf{(b)} Communication of one party's input to the other party (scenario (i) in the text, CPD). \textbf{(c)} Communication of one party's output to the other party (scenario (ii), COD). \textbf{(d)} Model with a message from one party to the other, possibly containing information about the input (scenario (iii), MCPD). The amount of communication can be measured by the entropy of the message.}
\label{fig.DAGs}
\end{figure}

\subsection{Causal models and correlations}

We focus on bipartite Bell scenarios where two spatially separated parties, Alice and Bob, receive inputs $x$, $y$ respectively, and produce outputs $a$, $b$. This corresponds e.g.~to a situation where Alice and Bob perform measurements on a shared physical system (such as a quantum state) with measurement settings $x$, $y$ and get outcomes $a$, $b$. The data from the experiment is encoded in the conditional probability distribution $p(a,b \vert x,y)$.

One can attempt to explain a given experiment in terms of a classical causal model. To do so, random variables are associated to each of the observed quantities ($A$ with $a$ etc.). In addition, to explain correlations in the observed data, common causes in the form of additional, unobserved variables affecting the observed quantities may be introduced. \figref{fig.DAGs}(a) shows a causal model for a bipartite Bell scenario without communication in which correlations in the outputs $A$, $B$ are mediated by an unobserved variable $\Lambda$. Not all $p(a,b \vert x,y)$ will be compatible with a given model. For example, \figref{fig.DAGs}(a) is a local hidden variable (LHV) model, and implies that $p(a,b \vert x,y)$ must be local.

In general, a causal model can be represented by a directed acyclic graph (DAG), where each of the nodes represent random variables and the directed edges connecting them signify causal influence. Given a DAG with variables $\left( V_1,\dots,V_n \right)$, any probability distribution compatible with this DAG can be decomposed as
\begin{equation}
\label{eq.markov}
p(v_1,\dots,v_n)= \prod_{i=1}^{n} p(v_i \vert pa_i),
\end{equation}
where $pa_i$ stands for the set of graph-theoretical parents of (nodes with an edge pointing to) the $i$-th node. Decomposition \eqref{eq.markov} is equivalent to each variable being independent of all its non-descendants given its parents \cite{Pearlbook}. Each variable $V_i$ can further be taken to be a deterministic function of its parents and some auxiliary, local noise term $U_i$. We then have that $v_i=f_i(pa_i,u_i)$.

For the LHV model \figref{fig.DAGs}(a), the condition \eqref{eq.markov} implies that $p(\lambda,x,y,a,b)= p(\lambda)p(x)p(y) p(a \vert x,\lambda) p(b \vert y,\lambda)$, from which it follows that
\begin{equation}
\label{eq.LHV}
p(a,b \vert x,y)= \sum_{\lambda} p(\lambda) p(a \vert x,\lambda) p(b \vert y,\lambda).
\end{equation}
This decomposition is the usual way in which the bipartite LHV model is defined. It can be understood to follow from the assumptions of realism, stating that measurement outcomes have well-defined values even if the measurement is not performed, and local causality, meaning that the local observations of Alice (the marginal distribution) cannot have a direct causal dependence on any actions or measurement results of Bob (that is, $p(a \vert x,y,b,\lambda)=p(a \vert x,\lambda)$) and vice versa \footnote{The idea of measurement independence is implicit in the discussion of Bell's theorem. That is, the experiments can choose which property of a system to measure independently of how the system has been prepared. See \cite{Chaves2015} for further details}.

Bell observed, that measurements on quantum entangled states may be incompatible with any LHV model. Imposing the same causal structure as in \figref{fig.DAGs}(a) but replacing the classical variable $\Lambda$ by a quantum state $\rho$, the conditional probabilities are given by the Born rule
\begin{equation}
\label{eq.Quantum}
p(a,b \vert x,y)= \mathrm{tr} \left( \rho \left( M^{x}_{a} \otimes M^{y}_{b}  \right)  \right),
\end{equation}
where $M^{x}_{a}$ and $M^{y}_{b}$ are measurement operators. For certain entangled $\rho$ and appropriate measurements, \eqref{eq.Quantum} does not admit any decomposition of the form \eqref{eq.LHV}.

Both local \eqref{eq.LHV} and quantum \eqref{eq.Quantum} correlations respect no signalling, which can be defined formally as the condition
\begin{equation}
\label{eq.NS}
\begin{split}
p(a\vert x) & = \sum_{b} p(a,b \vert x,y) \,\, \forall \, y \\
p(b\vert y) & = \sum_{a} p(a,b \vert x,y) \,\, \forall \, x .
\end{split}
\end{equation}
As pointed out by Popescu and Rohrlich \cite{Popescu1994}, and also studied by Tsirelson \cite{Tsirelson1993}, there are no-signalling distributions that have no quantum realisation. Denoting the sets of distributions $p(a,b|x,y)$ which admit a classical LHV model \eqref{eq.LHV}, a quantum realisation \eqref{eq.Quantum}, or respect no-signalling \eqref{eq.NS} by $\mathcal{C}$, $\mathcal{Q}$, and $\mathcal{NS}$ respectively, since any LHV model can be formulated using a separable quantum state, and since any quantum realisation respects no signalling, we then have the strict inclusions $\mathcal{C} \subsetneq \mathcal{Q}  \subsetneq \mathcal{NS}$. The main aim of this paper is to investigate how to simulate one set of correlations with another by allowing some communication. Specifically, how much communication do we have to add to the set of classical correlations $\mathcal{C}$ in order to reproduce points in $\mathcal{Q}$ or $\mathcal{NS}$? We also touch upon the question of how much communication must be added to $\mathcal{Q}$ to simulate $\mathcal{NS}$.

Before moving on, we note that, for finite numbers of inputs and outputs, $\mathcal{C}$ and $\mathcal{C}$ augmented by communication are polytopes, which facilitates testing whether a given distribution belongs to the sets. To be specific, consider the LHV model \figref{fig.DAGs}(a) with $n_a$, $n_b$ inputs for Alice and Bob, and $o_a$, $o_b$ outputs (i.e.~$x$ can take $n_a$ values etc.). Without loss of generality, $a$ can be taken to be a deterministic function of $x$ and $\lambda$, that is, $a=f_a(x,\lambda)$, and similarly $b = f_b(b,\lambda)$. Since there are $o_a^{n_a}$ functions mapping $x\rightarrow a$ and $o_b^{n_b}$ functions $y\rightarrow b$, it is then sufficient to consider $o_a^{n_a}o_b^{n_b}$ values for $\lambda$. Each value of $\lambda$ labels a choice of $f_a$, $f_b$, i.e.~a deterministic strategy for the outputs given the inputs. A distribution is compatible with the LHV model if and only if it can be decomposed in terms of these deterministic strategies. Collecting the probabilities $p(a,b|x,y)$ into a vector $\vec{p}$, the probabilities $p(\Lambda=\lambda)$ into a vector $\vec{q}$, and defining a matrix $T$ with elements $\delta_{a,f_a(x,\lambda)} \delta_{b,f_b(y,\lambda)}$, the decomposition becomes
\begin{equation}
\vec{p}= T \vec{q} .
\end{equation}
Checking the existence of such a decomposition for a given $\vec{p}$ can be cast as a simple linear program \cite{boyd_convex_2009,Chaves2015}. Alternatively, one can explicitly derive all the corresponding linear constraints. These define a convex polytope with finitely many extremal points corresponding to the deterministic strategies \cite{Pitowsky1989,Pitowsky1991}. The facets of the polytope are the Bell inequalities.

The no-signalling constraints \eqref{eq.NS} are also linear and $\mathcal{NS}$ is also a convex polytope. The quantum set $\mathcal{Q}$, on the other hand, is not a polytope and in general deciding if a given point lies in $\mathcal{Q}$ is a much harder problem, with the best known method consisting of a hierarchy of semi-definite programs \cite{Navascues2007}. This is a very active area of research \cite{Van1999nonlocality,Pawlowski2009,Navascues2010glance,Navascues2014almost,Fritz2013local,Sainz2014,Chaves2015information,ChavesBudroni2016} but not the focus of the present paper.

\subsection{Causal models for relaxations of locality}

The DAG representation of causal models provides a straightforward way to devise new scenarios where the assumption of local causality is explicitly relaxed. The basic rule consists in allowing arrows from one party to the other in such a way that no causal loops are generated. All the causal models that we consider here are shown in \figref{fig.DAGs}. We can identify three fundamentally different classes of locality relaxations.

In the first class, the input of one party has some direct causal influence over the outcomes of the other, as in \figref{fig.DAGs}(b). This DAG implies the decomposition
\begin{equation}
\label{eq.Xcomm}
p(a,b \vert x,y)= \sum_{\lambda} p(\lambda) p(a \vert x,\lambda) p(b \vert x,y,\lambda).
\end{equation}
Following the nomenclature in the literature \cite{Ringbauer2016}, we will refer to this sort of models as \emph{causal parameter-dependent} (CPD) models. Within this class, we can define one-way CPD models, corresponding to \figref{fig.DAGs}(b) and the equivalent with the roles of the parties reversed, and two-way CPD models, corresponding to \figref{fig.DAGs}(b) with an additional arrow from $Y$ to $A$. In the one-way models, $p(b \vert x,y,\lambda) \neq p(b \vert y,\lambda)$ or $p(a \vert x,y,\lambda) \neq p(a \vert x,\lambda)$ but not both simultaneously. In the two-way model, causal parameter independence is broken on both sides.

In the second class of models, the output of one party has some direct causal influence over the output of the other, as in \figref{fig.DAGs}(c). This implies
\begin{equation}
\label{eq.Acomm}
p(a,b \vert x,y)= \sum_{\lambda} p(\lambda) p(a \vert x,\lambda) p(b \vert a,y,\lambda).
\end{equation}
We will refer to this class as \emph{causal outcome-dependent} (COD) models \cite{Ringbauer2016}. Notice that in this case, we can only define a one-way class, as a two-way class where $A \rightarrow B$ and $B \rightarrow A$ would define a cycle in the graph where the variable $A$ or $B$ it is its own cause and therefore cannot be assigned any causal interpretation. However, in order to maintain the symmetry between the two parties we can allow the convex mixture of $A \rightarrow B$ and $B \rightarrow A$, leading to
\begin{equation}
\begin{split}
\label{eq.ABcommconvex}
p(a,b \vert x,y)= & \sum_{\lambda} p(\lambda) p(a \vert x,\lambda) p(b \vert a,y,\lambda) \\
& + \sum_{\mu} p(\mu) p(a \vert b,x,\mu) p(b \vert y,\mu) ,
\end{split}
\end{equation}
where $\sum_\lambda p(\lambda) + \sum_\mu p(\mu) = 1$.

In the third class we consider, causal parameter independence is relaxed via an intermediate variable representing a message, as in \figref{fig.DAGs}(d). We will refer to this model as \emph{message causal parameter dependent} (MCPD). Again, both one-way and two-way classes are possible. Considering e.g.~the one-way class $M \rightarrow B$, we have
\begin{equation}
\label{eq.Mcomm}
p(a,b \vert x,y) = \sum_{\lambda,m} p(\lambda)p(m \vert x,\lambda) p(a \vert x,\lambda) p(b \vert m,y,\lambda),
\end{equation}
This class of models allows additional control over what information is transmitted between the parties, for example by bounding the size of the message.

We remark that other classes of causal models are possible. For models with a message, we could consider the cases where the message is a function of the output (a message-outcome-dependent model) or of both the input and output (a message-parameter/outcome-dependent model). We will refrain from a more detailed analysis of these scenarios for two main reasons. First, as we will see, COD models are in many cases already insufficient to reproduce quantum or general no-signalling correlations. Thus, there is not much reason to consider the intermediate situation where only partial information about the outcome is communicated via a message. Second, for all the cases we have considered, allowing the message to depend on both the output and input does not provide any advantage over the case where it only depends on the input.

In each model \figref{fig.DAGs}(b)-(d), locality relaxation is achieved via a new causal link (with respect to \figref{fig.DAGs}(a)) influencing one of the outputs. In principle one could also consider new causal links influencing the inputs, e.g.~$X \rightarrow Y$ or $A \rightarrow Y$. However, since we are interested in simulating no-signalling distributions obtained by local measurements on a joint system, $p(x, y) = p(x)p(y)$ and thus a link of the type $X \rightarrow Y$ would be irrelevant. A link of the type $A \rightarrow Y$ means that $Y$ can be correlated with the unobserved variable $\Lambda$, implying some degree of measurement dependence. Measurement-dependent models are interesting and have also received lots of attention as a possible way to simulate nonlocal correlations \cite{Hall2010,Barrett2011,Chaves2015}. However, we will not consider them here.

\subsection{Quantifying locality relaxations}
In the following we introduce measures of the locality relaxations for the models in \figref{fig.DAGs}.

\textit{Direct causal influence.--} We start by considering the CPD model. Though it defines a valid model when the strength of the causal link $X\rightarrow B$ is arbitrary, it is trivial in the sense that the decomposition \eqref{eq.Xcomm} can reproduce all no-signalling distributions \cite{Bacon2003}. To simulate a distribution in $\mathcal{NS}$ via this model one proceed as follows. Alice and Bob share a distribution $p(\lambda)$ such that $p(a\vert x) = \sum_{\lambda} p(\lambda) p(a\vert x,\lambda)$ where $p(a\vert x,\lambda)$ is deterministic. By knowing $x$ and $\lambda$, Bob then unambiguously identifies $a$. Since any no-signalling distribution can be rewritten as $p(a,b \vert x,y) = p(b\vert x,y,a) p(a \vert x)$ it can then be simulated since Bob can locally generate $p(b\vert x,y,a)$.

The CPD model becomes non-trivial when the strength of the link $X\rightarrow B$ is limited. One natural way to define this strength, widely employed in the field of causal inference \cite{Pearlbook,Janzing2013}, is via the concept of \emph{intervention}. An intervention is the act of forcing a given variable $V_i$ to take on a specific value $v^{\prime}_i$. We denote it by $do(v^{\prime}_i)$. This erases the original mechanism $f_i(pa_i)$ defining the value of the random variable and introduces a new mechanism which sets $v_i$ to $v^{\prime}_i$ while keeping all other functions $f_j$ for $j \neq i$ unchanged. This changes the decomposition \eqref{eq.markov} to \footnote{{As noted in \cite{Chaves2015} the $do$-operation is defined relative to a causal model as encoded in the DAG, and in particular, $p(y \vert do(x))$ is the same as the usual conditional probability $p(y \vert x)$ only if the sets of parents $PA_X$ and $PA_Y$ are disjoint.}}
\begin{equation}
\label{markov_trunc}
	p(v_1,\dots,v_n \vert do(v^{\prime}_i))   =\left\{
\begin{array}{ll}
\prod_{j \neq i}^n p (v_j | \mathrm{pa}_{j} ) & \text{ if } v_i=v^{\prime}_i,\\
0 & \text{ otherwise.}%
\end{array}
\right.
\end{equation}
With the help of interventions, in \cite{Chaves2015} the \emph{direct causal influence} $\mathcal{C}_{X \rightarrow B}$ from $X$ to $B$ for the model in \figref{fig.DAGs}(b) was defined as
\begin{equation}
\label{meas_causal}
\mathcal{C}_{X \rightarrow B}= \sup_{b,y,x,x^{\prime}} \sum_{\lambda} p(\lambda) \vert p(b\vert do(x), y, \lambda)-p(b\vert do(x^{\prime}), y,\lambda )\vert.
\end{equation}
It is the maximum shift in the distribution of $B$ caused by interventions in $X$, averaged over the unobserved variable $\Lambda$. Similarly, one can define $\mathcal{C}_{A \rightarrow B}$ for the COD model in \figref{fig.DAGs}(d). This measure is strictly larger than zero for any underlying causal influence, as opposed to variations of it, such as the widely used \emph{average causal effect} that can be null even in the presence of causal influences \cite{Janzing2013}.

\textit{Communication entropy.--} For the MCPD model, we will focus on the average communication needed to simulate a given distribution (other measures can be considered and we refer the reader to \cite{Pironio2003} for a detailed discussion). We can categorize each of the deterministic strategies in \eqref{eq.Mcomm} according to the total number of bits required for the messages. For example, a strategy such that $p(m \vert x,\lambda)= \delta_{m,0}$ requires $0$ bits of information while $p(m \vert x,\lambda)= \delta_{m,x}$ (with $x=0,\dots,n_a-1$) requires a message with $\log_2 n_a$ bits. Given $\vec{p}$ compatible with \eqref{eq.Mcomm}, the average communication $\overline{C}(\vec{p})$ required is given by
\begin{equation}
\overline{C}(\vec{p})=\sum_{\lambda} p(\lambda) m_{\lambda},
\end{equation}
where $m_{\lambda}$ is the number of bits required by the fixed strategy $\lambda$.

Another option is the Shannon entropy of the message \cite{Chaves2015entropy}, which is closely related to, but different from, the average communication and which, to our knowledge, has not yet been considered. Given a distribution $\vec{p}$ that can be simulated by the model \eqref{eq.Mcomm}, the entropy of the message is given by $H(M)=-\sum_{m}p(m)\log_2 p(m)$.

The computation of each of the measures above can be formulated in terms of linear programs. For the minimisation of the measures $\mathcal{C}_{A \rightarrow B}$, and $\mathcal{C}_{X \rightarrow B}$ for given observed data, or a given Bell inequality violation, this was shown in \cite{Chaves2015} (and accompanying supplementary information). The minimisation problem for the communication entropy was discussed in the supplementary information of \cite{Chaves2015entropy}. The observed distribution $p(a,b\vert x,y)$ implies restrictions on the distribution $p(m)$ of the message, again defining a polytope. Because of convexity of the entropy, it is only necessary to consider the extremal points of this polytope, which considerably reduces the computational complexity. The general method for finding the polytope is given in \cite{Chaves2015entropy} and requires running a sequence of linear programs.

Finally, we note that rather than computing the minimal relaxation of locality (according to some measure) required to simulate a given distribution, we may be interested in fixing a certain communication scenario and derive the corresponding Bell inequalities \cite{Bacon2003,Maxwell2014,Chaves2015}. As detailed above, the CPD model can simulate all no-signalling correlations and thus no Bell inequalities in this scenario will be violated without signalling. The situation, however, is different for the COD and MCPD models.

\section{Communication of inputs}
\label{sec.locrelax}

In the supplementary information of \citep{Chaves2015}, it was shown that for the CPD model \figref{fig.DAGs}(b) with two inputs and two outputs per party, one has 
\begin{equation}
\min \mathcal{C}_{X \rightarrow B}= \max \left[I_{CHSH},0 \right]
\end{equation}
where
\begin{eqnarray}
I_{CHSH}=& & p(00\vert 00)+p(00\vert 01)+p(00\vert 10) \\\nonumber
& & -p(00\vert 11)-p_{A}(0\vert 0)-p_{B}(0\vert 0)\leq 0,
\end{eqnarray}
is an equivalent form of the famous Clauser-Horne-Shimony-Holt (CHSH) Bell inequality \cite{Clauser1969}.

\begin{figure}[!t]
\includegraphics[width=0.93\columnwidth]{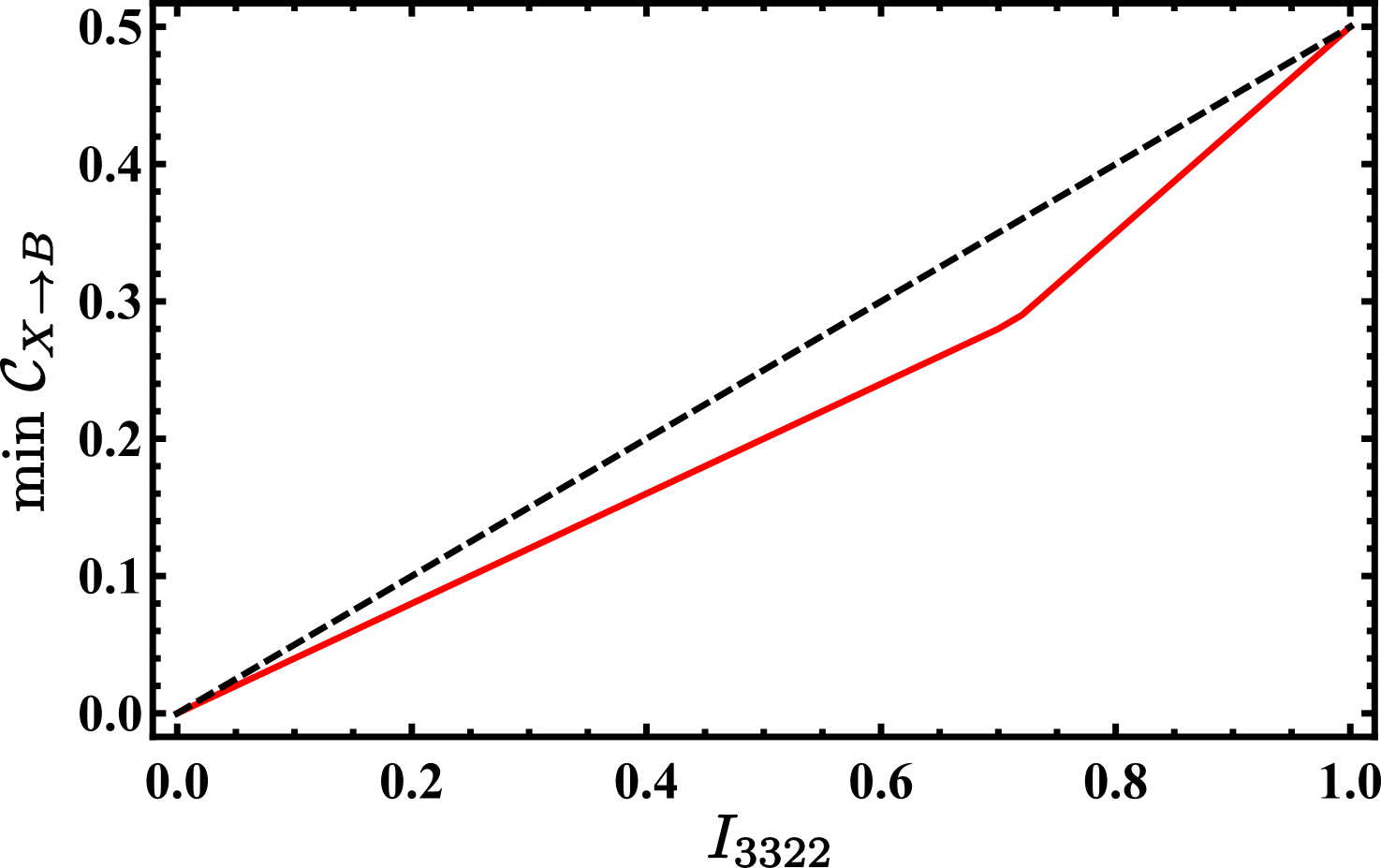}
\caption{The value of $\min \mathcal{C}_{X \rightarrow B}$ as function of the $I_{3322}$ value. The black dashed curve is for the case where the full probability distribution \eqref{eq:I3322_distribution} is taken into account. The red solid curve is for the case where only the value of $I_{3322}$, non-signalling, and normalization constraints are imposed. As opposed to the CHSH scenario, fixing the $I3322$ value gives only a lower bound on the minimum causal influence required to explain given nonlocal correlations.
}
\label{fig:I3322plot}
\end{figure}

Given this causal interpretation of the CHSH inequality, one can wonder whether similar results hold for other Bell inequalities, e.g.~the inequality $I_{3322}$ for a scenario with three inputs and two outputs per party given by \cite{Collins2004}
\begin{align}
I_{3322} & = p(00\vert 00)+p(00\vert 01)+p(00\vert 02)+p(00\vert 10) \nonumber \\
& + p(00\vert 11)-p(00\vert 12)+p(00\vert 20)-p(00\vert 21) \nonumber \\
& - 2p_{A}(0\vert 0)-p_{A}(0\vert 1)-p_{B}(0\vert 0) \leq 0.
\end{align}
It turns out, as we now show, that the $I_{3322}$ inequality only provides a lower bound on the value of $\mathcal{C}_{X \rightarrow B}$ required for simulating nonlocal distributions. This is illustrated in \figref{fig:I3322plot}. We consider the particular distribution
\begin{equation}
p(a,b \vert x,y)= v p_{\text{PR}}+(1-v) p_{\text{W}},
\label{eq:I3322_distribution}
\end{equation}
where
\begin{equation}
p_{\text{PR}}\left(  a,b | x,y \right)  =\left\{
\begin{array}{ll}
1/2 & \text{if } a+ b=1\mod 2 \text{, } x+y=3,   \\
1/2 & \text{if } a+ b=0 \mod 2 \text{, } x+y \neq 3,   \\
0 & \text{otherwise,}%
\end{array}
\right.
\end{equation}
corresponds to a generalization of the Popescu-Rohrlich (PR) box \cite{Popescu1994} and maximally violates the $I_{3322}$-inequality (achieving $I_{3322} = 1$), and
\begin{equation}
p_{\text{W}}\left(  a,b | x,y \right)  =1/4
\end{equation}
denotes the uniform distribution (for which $I_{3322} = -1$). The distribution \eqref{eq:I3322_distribution} gives $I_{3322}=2v-1$.
We numerically see that, taking into account the full probability distribution,
\begin{equation}
\mathcal{C}_{X \rightarrow B}=\max \left[0,(2v-1)/2\right]=\max \left[0,I_{3322}/2\right] .
\end{equation}
However, if we instead only impose a fixed value of $I_{3322}$ (plus no-signalling and normalization constraints) we get
\begin{equation}
\min \mathcal{C}_{X \rightarrow B}
=
\left\{
\begin{array}{ll}
0 &  I_{3322} \leq 0, \\
(2/5)*I_{3322} & 0 \leq I_{3322} \leq 0.714,   \\
(1/4)*(3I_{3322}-1)& 0.714 \leq I_{3322} \leq 1.
\end{array}
\right.
\end{equation}
This shows that the requirements to simulate different distributions achieving the same value of $I_{3322}$ may be quite different. Moreover, this result highlights another nice aspect of the framework of \cite{Chaves2015}. Unlike the results in \cite{Hall2010,Hall2010a,Hall2011,Banik2013,Rai2012,Koh2012,Biswajit2013}, it can take into account the full probability distribution, not just the value of a specific Bell inequality.

\section{Communication of outputs}
\label{sec.outcomes}

Similarly to the case of communication of inputs, for communication of outputs as in \figref{fig.DAGs}(c), it has been proven that \cite{Chaves2015}
\begin{equation}
\label{CAB_CHSH_app}
\min \mathcal{C}_{A \rightarrow B}= \max \left[ 0,I_{CHSH} \right].
\end{equation}
This implies that such a model (where one party communicates the output) is capable of simulating any nonlocal distribution in the CHSH scenario.

Interestingly however, in a scenario with three inputs and two outputs per party, COD models are not enough to reproduce nonlocal correlations \cite{Chaves2015}. One of the inequalities characterizing the model \eqref{eq.Acomm}, corresponding to \figref{fig.DAGs}(c), is given by
\begin{eqnarray}
\label{chained33}
I_{A\rightarrow B}= & &\langle A_0 B_0 \rangle  - \langle A_0 B_2 \rangle
- \langle A_1 B_1 \rangle  \\ \nonumber
& &+ \langle A_1 B_2 \rangle-\langle A_2 B_0 \rangle + \langle A_2 B_1 \rangle  \leq 4,
\end{eqnarray}
where
\begin{equation}
\langle A_x B_y \rangle \sum_{a,b} (-1)^{a+b} p(a,b\vert x,y) .
\end{equation}
This inequality is invariant under party permutations, meaning that the same inequality remains valid if we replace $A \rightarrow B$ by $B \rightarrow A$ or consider convex combinations of the two. The inequality \eqref{chained33} can be violated by local measurements on quantum states. This shows that relaxing some of the assumptions in Bell's theorem is not necessarily enough to causally explain quantum correlations, a result that has recently been put to experimental test \cite{Ringbauer2016}.

\subsection{Generalization to more inputs}

We note that, because of the locality relaxation, not all inequalities that are equivalent in an LHV scenario will also be equivalent for COD models. In particular, in an LHV scenario, inequality \eqref{chained33} is equivalent (up to relabellings of parties, inputs, and outputs) to the chained Bell inequality proposed in \cite{Braunstein1990} which, in its canonical form, is given by
\begin{eqnarray}
\label{chained33can}
I_{\mathrm{Chained}}= & &\langle A_0 B_0 \rangle  - \langle A_0 B_2 \rangle
+ \langle A_1 B_0 \rangle \\ \nonumber
& &+\langle A_1 B_1 \rangle+\langle A_2 B_1 \rangle + \langle A_2 B_2 \rangle  \leq 4,
\end{eqnarray}
However, this canonical form does not define a valid inequality for \figref{fig.DAGs}(c). In fact, the COD model $A \rightarrow B$ can violate \eqref{chained33can} up to its algebraic maximum, using the following strategy. Alice uses a protocol where if $x = 1,2$ she outputs $a = 0$ and if  $x = 0$ she outputs $a = 1$. Bob uses a protocol where if $a = 0$ then $b = 0$, if $a = 1 \text { and } y =0$ then $b= 1$ and if $a = 1 \text{ and } y =2$ then $b= 0$. This way we get $6$ for the left-hand-side of \eqref{chained33can}.

In order to turn \eqref{chained33can} into \eqref{chained33} we need to apply the following transformations: $A_0 \rightarrow -A_0$, $A_2 \rightarrow -A_2$, $B_1 \rightarrow -B_1$, $B_0 \leftrightarrow B_2$. With this relabelling, any protocol of the type above will cease to work, and the inequality becomes valid for the COD model. For a qualitative understanding of the reason, consider that Alice applies the strategy above on her side. Bob can still make the term $\mean{A_0B_0}-\mean{A_0B_2}+\mean{A_1B_2}-\mean{A_2B_0}$ equal to $4$ but now, because the variable $a$ cannot distinguish between $x=1$ and $x=2$, the term $-\mean{A_1B_1}+\mean{A_2B_1}$ will always be equal to $0$. Roughly speaking, the symmetry transformations cannot be matched by the arrow $A \rightarrow B$.

Since the chained Bell inequality \eqref{chained33can} can be defined for an arbitrary number of inputs, these results naturally lead to the question whether some of its symmetries define valid inequalities for the COD model \figref{fig.DAGs}(c) with any number of inputs. In order to tackle this question, for varying numbers of inputs we have generated all the symmetries of the chained inequality as well as all the vertices of the polytope corresponding to \figref{fig.DAGs}(c) (the deterministic strategies). Evaluating the inequalities over all vertices, we check whether they can be violated by the COD model. For even numbers of inputs (up to 8), we find that all symmetries can be violated. However, for 3,5,7, and 9 inputs, we find that the following inequality can be made valid for the COD model
\begin{equation}
\label{eq.validchained}
\begin{split}
[a_1-b_1]+[b_1-a_2]+[a_2-b_2]+\cdots+[a_n-b_n] \\
+ [b_n-a_1-1] \geq 1 .
\end{split}
\end{equation}
Here, we follow the notation of \cite{Brunner2014} where $[a_x-b_y] = \sum_{j=0}^1 j p(a-b=j \,\text{mod}\, 2|x,y)$, and $a,b\in\{0,1\}$ and $x,y\in\{1,\ldots,n\}$. For the inequality to hold for the COD model, Alice's outputs must be relabelled for every second input. That is $a=0 \leftrightarrow a=1$ for even $x$. With this relabelling, we conjecture that \eqref{eq.validchained} holds for any odd number of inputs. Since this inequalities can be violated by local measurements on entangled states, this generalizes the result in \cite{Chaves2015} and shows that COD models are insufficient to reproduce quantum correlations for a larger number of inputs as well.

Notice that we have restricted our attention to the case of binary outcomes. Understanding the validity of generalizations of the chained inequality for more outcomes \cite{Barrett2006} is another interesting open question.

\subsection{More outputs than inputs}

As noted above, communication of the input (see \figref{fig.DAGs}(b)) is sufficient to reproduce any no-signalling distribution. When the number of outputs of Alice is equal to or larger than the number of inputs, one might then intuitively expect that the COD model \figref{fig.DAGs}(c) can also reproduce any no-signalling distribution, because the output space is large enough to encode Alice's input and thus communicate it to Bob. However, if a given distribution must reproduced, Alice's output cannot be chosen completely freely, and thus it is not a priory clear whether this intuition holds. In the following we prove that it holds for a restricted class of scenarios, and demonstrate that it is, however, not generally valid.

\subsubsection{Uniform marginals}

We first consider scenarios with $n$ inputs and $o$ outputs for Alice, and distributions with uniform marginals, that is $p(a|x) = 1/o$ for all $a$ and $x$. If $o \geq n$, then there exist injective functions from the set of inputs to the set of outputs. We let $\lambda$ label all such injective functions $f_\lambda$. There are $o!/(o-n)!$ of them. Since Alice's marginal is uniform, we can then write
\begin{equation}
\frac{1}{o} = p(a|x) = \frac{(o-n)!}{o!} \sum_\lambda \delta_{a,f_\lambda(x)} .
\end{equation}
One can see that the equality holds because, for fixed $x$ and $a$, there are $(o-1)!/(o-n)!$ injective functions such that $f_\lambda(x)=a$. The distribution can now be simulated as follows. Alice and Bob share $\lambda$ which is uniformly random. Alice outputs $a = f_\lambda(x)$ and communicates $a$ to Bob. From the pair $(a,\lambda)$, the input $x$ is uniquely determined and hence known to Bob. Bob now outputs $b$ according to $p(b|a,x,y) = p(a,b|x,y)/p(a|x)$, which he can do locally (note that any local randomness needed by Bob can be absorbed in $\lambda$). The resulting statistics correctly reproduces $p(a,b|x,y)$ as desired.

Next, we note that the argument can also be adapted to the case where certain values of $a$ never occur (for any $x$). If the number of outcomes that do occur $k$ is still large enough, $k \geq n$, and the distribution is uniform on the outcomes that do occur, then we can replace $1/o$ above by $1/k$ and restrict $\lambda$ to injective functions from $x$ into the set of outcomes with non-zero probability. 

Finally, we note that, for the model \figref{fig.DAGs}(c) to reproduce all no-signalling distributions in a given scenario, it is necessary and sufficient that it can reproduce each of the vertices of the no-signalling polytope. Since the model allows for shared randomness, any convex combination of reproducible distributions is also reproducible. Any local vertex can be trivially reproduced, so it is sufficient to look at the nonlocal vertices.

In \cite{Barrett2005}, the authors identify all the nonlocal no-signalling vertices for bipartite scenarios with binary inputs and any number of outputs. These are exactly of the form where some outputs never occur independent of $x$ (but at least two do) and Alice's marginal is uniform on outputs with non-zero probability (see \cite{Barrett2005} Eq.~(12)). Hence, it follows that the COD model \figref{fig.DAGs}(c) reproduces all no-signalling distributions in these scenarios.

\subsubsection{Non-uniform marginals}

For non-uniform marginals it is not generally possible to reconstruct all no-signalling distributions, even when the output space is larger than the input space and all outputs occur with non-zero probability. 

This can be seen because the no-signalling polytope of a scenario with at least as many outputs as inputs may contain vertices that are effectively lifted from lower scenarios with less inputs than outputs and which cannot be reproduced by the COD model. Such vertices will have an output from Alice which never occurs (so that the effective number of outputs is less than the number of inputs). However, by mixing them with uniform noise, one can construct distributions where all outputs have non-zero probability but which are still not reproducible within the model. 

We give an explicit example for the scenario [(333)(32)] (using the standard notation $[(o^A_1,\ldots,o^A_{n_A}) (o^B_1,\ldots,o^B_{n_B})]$, where there $o^A_i$, $o^B_i$ are the number of outcomes for each input of Alice and Bob respectively). Using the polytope software PORTA, we find all the vertices of the no-signalling polytope in this scenario, and using linear programming we can then check for each vertex if it can be reproduced by the model \figref{fig.DAGs}(d). \tabref{tab.nsvtx} shows a vertex which cannot be reproduced. We see that the outcome '0' for Alice never occurs. One can check that, ignoring this outcome, the vertex is also valid for the scenario [(222)(32)] where indeed we would not expect it to be reproducible.

\begin{table}
\begin{tabular}{c | c || c c c | c c |}
$x\backslash y$ & & \multicolumn{3}{c|}{0} & \multicolumn{2}{c|}{1} \\
\hline
 & $a\backslash b$      &   0            &   1             &   2   &   0   &   1   \\
\hline
\hline
\multirow{3}{*}{0} & 0  &   0           &   0           &   0            &       0        &      0        \\
                   & 1  &   0           &   0           &   0            &       0        &      0        \\
                   & 2  &   0           & $\frac{1}{2}$ & $\frac{1}{2}$  & $\frac{1}{2}$  & $\frac{1}{2}$ \\
\hline 
\multirow{2}{*}{1} & 0  &   0           &   0           &   0            &       0        &      0        \\
                   & 1  &   0           &   0           & $\frac{1}{2}$  &       0        & $\frac{1}{2}$ \\
                   & 2  &   0           & $\frac{1}{2}$ &   0            & $\frac{1}{2}$  &      0        \\
\hline
\multirow{2}{*}{2} & 0  &   0           &   0           &   0            &       0        &      0        \\
                   & 1  &   0           &   0           & $\frac{1}{2}$  & $\frac{1}{2}$  &      0        \\
                   & 2  &   0           & $\frac{1}{2}$ &   0            &       0        & $\frac{1}{2}$ \\
\hline
\end{tabular}
\caption{A no-signalling vertex of the scenario [(333)(32)] which cannot be reproduced by the model \figref{fig.DAGs}(d). The table entries are $p(a,b|x,y)$.}\label{tab.nsvtx}
\end{table}

More interestingly, we also find that the vertex remains impossible to reproduce when mixed with noise. Denoting the vertex in \tabref{tab.nsvtx} $p_{vtx}$ and the uniform distribution $p_{id}(a,b|x,y) = 1/(o^A_x o^B_y)$, we find that the distribution $(1-\epsilon) p_{vtx} + \epsilon p_{id}$ cannot be reproduced for any $\epsilon \lesssim 0.39$.

Thus it is not sufficient in general to have at least as many outputs as inputs to reproduce all no-signalling distributions. For a number of scenarios though, all the no-signalling vertices can be reproduced by the model \figref{fig.DAGs}(c). In particular, we have tested that this is the case for the scenarios [(2 2)(3 3)], [(2 2)(2 2 2)], [(3 2)(2 2 2)], [(2 2)(3 3 3)], and [(3 2)(3 3 3)].

\section{Communication of a message}
\label{sec.message}

Here we consider scenarios with communication of a message as in \figref{fig.DAGs}(d), where the amount of communication may be limited.

\subsection{New inequalities}

In Ref.~\cite{Bacon2003}, all the Bell inequalities constraining the scenario with binary inputs and outputs, and at most one bit of communication, were derived. In Ref.~\cite{Maxwell2014} this was generalised to allow three inputs for one of the parties. These results relied on the computational characterization of the associated polytopes. However, given the exponential increase in complexity with the number of inputs, this is not viable for more complicated scenarios. In order to circumvent this problem and get constraints also for the case of three inputs for both parties, Ref.~\cite{Bacon2003} considered the set of Bell inequalities with full correlators only. That is, instead of considering the full probability distribution $p(a,b \vert x,y)$ one restricts attention to constraints on the level of $\mean{A_xB_y}= \sum_{a,b}(-1)^{a+b}p(a,b \vert x,y)$ (where $a,b\in\{0,1\}$). In this case, one gets two inequivalent classes of inequalities, one of them given by
\begin{equation}
\label{M33}
\sum_{x,y=0,1,2}M_{x,y} \mean{A_xB_y} \leq 1,
\end{equation}
with the coefficients $M_{x,y}$ defined by the matrix
\begin{equation}
M = \frac{1}{6}\left( \begin{array}{ccc}
1 & 1 & 1 \\
1 & 1 & -1 \\
1 & -1 & 0 \end{array} \right).
\end{equation}
This inequality cannot be violated by quantum mechanics but the left-hand side of it can reach the value of $8$ for general non-signalling correlations. Unfortunately, however, even considering such simplifications one cannot move beyond the known cases in this computational approach. Instead, we sketch an analytical derivation for the generalization of \eqref{M33}.

We start considering the generalization for the case of $n$ inputs and binary message. In this case, the analogue of \eqref{M33} holds with
\begin{equation}
\label{Imm_prob}
M^{n}_{2} = \frac{1}{3+n(n-1)/2}\left(
\begin{tabular}{cccccc}
  1 & 1 & $\cdots$ & 1 & 1 & 1\\
   1 & 1 & $\cdots$ & 1 & 1& -1 \\
  1 & 1 & $\cdots$ & 1 & -1 & 0\\
  $\vdots$ &  $\vdots$ & $\vdots$ & $\ddots$ & $\vdots$ & $\vdots$\\
   1 & -1 & 0 &  $\cdots$ & 0 & 0\\
\end{tabular}
\right).
\end{equation}
Note that the maximum algebraic value we can achieve in each row equals $n$ for the first and second rows and $n-i$ for each $i$'th remaining row. 

The reason for the factor $\frac{1}{3+n(n-1)/2}$ is simple. There are $n$ possible inputs but we can send a message with two symbols only. Let's say we send $m=0$ if $x=0$ and $m=1$ otherwise. In this case, Bob knows whenever $x=0$ but nothing else. If he receives the message $m=0$ he generates outcomes such that he is maximally correlated with Alice, that is, $\mean{A_0B_j}=1$ (the first row in the matrix). But if $m=1$ he does not know Alice's input. The best he can do is to maximize one of the rows in the matrix (it does not matter which one). For example, he can generate the maximum score for the second row. Thus, with a two-symbol message he can score $2n$ in the first two rows. In the remaining rows he uses the same strategy as in the second row and thus scores $\sum_{i=0}^{n-3}(n-i-3)$. So in total he scores $3+n(n-1)/2$ which lead us to the bound.

The same idea holds if we consider a message with more symbols. For general message dimension $d$ we can get perfect score in the first $d$ rows. In the remaining rows we score $\sum_{i=0}^{n-d-1}(n-i-d-1)$ and the bound in this case is given by $(n(n-1)+d+d^2)/2$. That is, the general inequality for $n$ inputs and a message with $d$ symbols is given by the analogue of \eqref{M33} holds with
\begin{equation}
\label{Imm_gen}
M^{n}_{d} = \frac{1}{(n(n-1)+d+d^2)/2}\left(
\begin{tabular}{cccccc}
  1 & 1 & $\cdots$ & 1 & 1 & 1\\
   1 & 1 & $\cdots$ & 1 & 1& -1 \\
  1 & 1 & $\cdots$ & 1 & -1 & 0\\
  $\vdots$ &  $\vdots$ & $\vdots$ & $\ddots$ & $\vdots$ & $\vdots$\\
   1 & -1 & 0 &  $\cdots$ & 0 & 0\\
\end{tabular}
\right).
\end{equation}

\subsection{Quantifying the entropy of the message}

Rather than deriving inequalities for the scenario \figref{fig.DAGs}(d), we now ask how much communication is required to reproduce a given value of a known Bell inequality for the LHV scenario \figref{fig.DAGs}(a). Specifically, we quantify the minimal entropy $H(m)$ of the message required for this model to explain the observations. In Ref.~\cite{Chaves2015} this question was addressed in the simplest case of binary inputs and outputs, and a binary message. It was found that the minimum Shannon entropy of the message was exactly equal to the binary entropy of the CHSH violation, $\min H(m) = \max[h(I_{CHSH}),0]$ (for an appropriate formulation of CHSH). The derivation of this result was particular to a binary message and could not easily be extended. Here, we adopt methods from Ref.~\cite{Chaves2015entropy}, which allow us to treat larger message alphabets as well as more inputs and outputs. We recover the result for the CHSH scenario, and treat scenarios with more inputs and outputs as well.

\begin{figure*}[t]
  \centering
  \includegraphics[width=\textwidth]{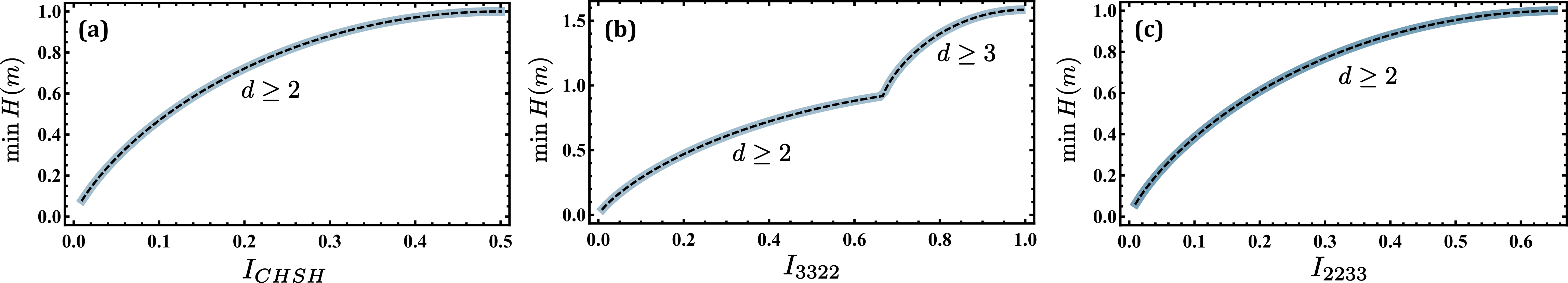}
  \caption{Minimal message entropy vs.~Bell parameter value. Blue, solid curves are computed by minimising the entropy via linear programming, for message dimensions $d=2,3,4$. Black, dashed curves are analytical functions of the Bell parameter. The message dimension required to reach different regions are indicated. \textbf{(a)} CHSH. The dashed curve is $h(I_{CHSH})$. \textbf{(b)} $I_{3322}$. The dashed curve is $h(I_{3322})$ for $I_{3322}\leq 2/3$ and $(2/3)h(3I_{3322}/2-1) + h(1/3)$ for $I_{3322} > 2/3$. \textbf{(c)} $I_{2233}$. The dashed curve is $h(3I_{2233}/4)$.}
  \label{fig.SvsHplots}
\end{figure*}

Specifically, we test scenarios with 2 inputs, 2 outputs, with 3 inputs, 2 outputs, and with 2 inputs, 3 ouputs, corresponding to the Bell inequalities CHSH, $I_{3322}$, and $I_{2233}$ respectively \cite{Collins2004}. In each case for message dimensions up to $d=4$. We use the following form of the inequalities, and we impose no-signalling
\begin{align}
I_{CHSH} = & -p(00|00)+p(00|01)+p(00|10) \nonumber \\
& -p(00|11)-p(01|00)-p(10|00) \leq 0 , \\
I_{3322} = & -2p(00|00)+p(00|02)+p(00|10) \nonumber \\
& +p(00|11)-p(00|12)+p(00|20) \nonumber \\
&-p(00|21)-p(01|00)-2p(10|00) \nonumber \\
&-p(10|01) \leq 0 , \\
I_{2233} = & -p(00|00)-p(01|00)+p(01|01) \nonumber \\
& +p(01|10)-p(01|11)-p(02|00) \nonumber \\
& -p(10|00)+p(10|01)+p(10|10) \nonumber \\
& -p(10|11)-2p(11|00)+p(11|01) \nonumber \\
& +p(11|10)-p(11|11)-p(12|00) \nonumber \\
& -p(20|00)-p(21|00) \leq 0 .
\end{align}

The results are shown in \figref{fig.SvsHplots}. We plot the minimal message entropy vs.~the Bell parameter value. For CHSH we find that 
\begin{equation}
\min H(m) = h(I_{CHSH})
\end{equation}
for any dimension of the message ($d=2,3,4$).

For $I_{3322}$ we find that for a violation up to $I_{3322}=2/3$, which can be reached with $d=2$ one has
\begin{equation}
\min H(m) = h(I_{3322}/2) ,
\end{equation}
while in the region from $2/3$ to 1 (the no-signalling bound), which can be reached with $d=3$, one has a different relation
\begin{equation}
\min H(m) = \frac{2}{3} h(\frac{3}{2}I_{3322}-1) + h(\frac{1}{3}) .
\end{equation}
In the region reachable by $d=2$, increasing the message dimension to 3 or 4 brings no advantage. Similarly, in the region reachable by $d=3$, increasing to 4 brings no advantage.

For $I_{2233}$ a maximal violation of $I_{2233} = 2/3$, which is the no-signalling bound, can be reached for $d=2,3,4$. We find the relation
\begin{equation}
\min H(m) = h(\frac{3}{4} I_{2233}) .
\end{equation}
Again, increasing the message dimension beyond 2 brings no advantage.

\section{Augmenting quantum mechanics with communication}
\label{sec.augmenting}

In the previous sections we have considered how communication can be used to simulate nonlocal correlations via classical models. A related question is how communication can be used together with quantum correlations in order to achieve even stronger correlations, a scenario that can be analyzed from two complementary perspectives.

First, from a causal perspective, Bell's theorem can be understood as comparing the classical and quantum descriptions when imposing a given causal structure on an experiment. In the usual Bell scenario, because there is no communication between the parties (enforced e.g. by space-like separation), all their correlations must be mediated by a common source (a classical hidden variable or an entangled quantum state). In this new scenario, again we are interested in comparing classical and quantum resources, the difference being that now we augment the underlying (classical or quantum) causal models with some limited amount of communication.  

Second, it is interesting to understand how much communication we have to add to quantum correlations in order to be able to reproduce general no-signalling (postquantum) correlations.

We discuss a particular case of this general question, focusing on a scenario with three inputs and binary outputs. In this case, we have seen that classical models with one bit of communication are bounded by the inequality given by
\begin{eqnarray}
\label{M332}
& & \mean{A_0B_0}+\mean{A_0B_1}+\mean{A_0B_2}+
\mean{A_1B_0} \\ \nonumber
& & +\mean{A_1B_1}-\mean{A_1B_2}+
\mean{A_2B_0}-\mean{A_2B_1} \leq 6
\end{eqnarray}
Quantum correlations (obtained by local measurements and no communication) are bounded by the same inequality. However, more general no-signalling correlations can violate the inequality, showing that, at least for classical models, one bit of communication is not enough to simulate all nonlocal correlations in this scenario. However, as we show next, if we augment quantum correlations with (limited) communication, then inequality \eqref{M332} can also be violated by proper local measurements on a entangled state.

The protocol proceed as follows. Alice and Bob share a maximally entangled state $\ket{\Psi^{+}}=(\ket{00}+\ket{11})/\sqrt{2})$ and Alice performs local measurements given by $A_0=A_1=Z$, $A_2=X$ (where $X$ and $Z$ are the Pauli operators). If Alice receives $x=0$, she sends a message $m=0$, if $x=1$ or $x=2$ she sends $m=1$. If we assume that all the inputs are equally likely, the message has less than one bit of information, since $H(m) \approx 0.92$. Notice that \eqref{M332} includes the terms $\mean{A_1B_0}+\mean{A_1B_1}+
\mean{A_2B_0}-\mean{A_2B_1}-\mean{A_1B_2}$ corresponding to the CHSH operator plus one extra term ($-\mean{A_1B_2}$). If Bob receives $m=1$, he measures $B_0=(Z+X)/\sqrt{2}$, $B_1=(Z-X)/\sqrt{2}$ and $B_2=-Z$ achieving $2 \sqrt{2}+1$ for this combination of 5 terms. For $m=0$, Bob measures $B_0=B_1=B_2=Z$ achieving the maximal possible value of $3$ for the remaining terms $\mean{A_0B_0}+\mean{A_0B_1}+\mean{A_0B_2}$ appearing in \eqref{M332}. This way, with less than one bit of communication we achieve the value of $4+2\sqrt{2}$, violating the inequality.

\section{Discussion and Outlook}
\label{sec.discussion}

In this paper we have studied Bell scenarios where the locality assumption is relaxed to allow for some amount of communication. We have considered sets of correlations $\mathcal{C}$, $\mathcal{Q}$, and $\mathcal{NS}$ which can be obtained from classical causal models, quantum causal models, and general no-signalling models respectively, and we have asked how much communication one needs to add to $\mathcal{C}$ in order to reproduce points in $\mathcal{Q}$ or $\mathcal{NS}$? Focusing on bipartite scenarios, we have considered communication of inputs, of outputs, and of a message, and have derived several generalizations of previous results and analysed new models and measures of locality relaxations. 

For communication of inputs, we demonstrated that in general, fixing a given Bell inequality value provides only a lower bound on the minimal causal influence required to simulate no-signalling correlations.

For communication of outputs, we identified a family of inequalities bounding the set of classical correlations simulatable with communication for varying number of inputs. We further considered scenarios with at least as many outputs as inputs. For two inputs and two outputs, all no-signalling correlations can be simulated. One might think this could be a general feature since with more outputs than inputs it is in principle possible to encode the inputs in the outputs, and we know that (unrestricted) communication of the inputs enable simulation of all no-signalling correlations. We have demonstrated that this intuition holds true in the case where the marginals are uniform. However, perhaps surprisingly, we also have found that it does not generally hold for non-uniform marginals. We note that a non-loophole-free experimental test with communication of outcomes was recently implemented \cite{Ringbauer2016}. New inequalities in this setting (with more inputs and outputs) could potentially lead to relaxed detection efficiency thresholds for loophole-free tests. Further insight in this setting may also lead to stronger cryptographic protocols where an eavesdropper would be allowed some access to the parties outcomes. 

For communication of a message, we identified another family of full-correlator inequalities for arbitrary number of inputs bounding the set of classical correlations simulatable with a binary message. We also found simple relations between the minimal message entropy required for simulation, the the values of the CHSH, $I3322$, and $I2233$ Bell inequalities.

Finally, we introduced a new kind of scenario, asking how much communication must be added to $\mathcal{Q}$ to simulate $\mathcal{NS}$. That is, augmenting quantum correlations with classical communication. We believe that this kind of scenarios are interesting for several reasons. Bell scenarios with some sort of communication play an important role in communication complexity problems \cite{Buhrman2010} and this scenario may lead to useful generalizations. Communication also plays an important role in information-theoretic principles for quantum correlations, such as the celebrated information causality \cite{Pawlowski2009}. Thus, understanding how much communication we have to add to quantum mechanics in order to achieve stronger, post-quantum correlations could lead to new insights into the nature of quantum correlations themselves.

\begin{acknowledgments}
We acknowledge helpful discussions with S.~Pironio. JBB acknowledges support from the Swiss National Science Foundation. RC acknowledges financial support from the Brazilian ministries MEC and MCTIC, the FQXi Fund, the Excellence Initiative of the German Federal and State Governments (Grants ZUK 43 \& 81), the US Army Research Office under contracts W911NF-14-1-0098 and W911NF-14-1-0133 (Quantum Characterization, Verification, and Validation), the DFG (GRO 4334 \& SPP 1798).
\end{acknowledgments}

\bibliography{Bell_comm_bib}

\end{document}